\documentclass[AMA,STIX1COL]{WileyNJD-v2}
\usepackage{moreverb}
\usepackage{csquotes}
\usepackage{graphicx,color}
\usepackage{float}
\usepackage[caption=false]{subfig}
\usepackage{amsmath}
\usepackage{amsmath}
\usepackage{multirow}
\usepackage{dsfont}
\usepackage[shortlabels]{enumitem}
\usepackage{epstopdf}

\newcommand\BibTeX{{\rmfamily B\kern-.05em \textsc{i\kern-.025em b}\kern-.08em
T\kern-.1667em\lower.7ex\hbox{E}\kern-.125emX}}
\articletype{ORIGINAL ARTICLE}%
\received{<day> <Month>, <year>}
\revised{<day> <Month>, <year>}
\accepted{<day> <Month>, <year>}

\begin{document}

\title{On the Klimontovich description of complex (dusty) plasmas}
\author{Panagiotis Tolias}
\authormark{P. Tolias}
\address{\orgdiv{Space and Plasma Physics}, \orgname{Royal Institute of Technology (KTH)}, \orgaddress{\state{Stockholm}, \country{Sweden}}}
\corres{Panagiotis Tolias, Space and Plasma Physics, Royal Institute of Technology, Stockholm SE-100 44, Sweden. \email{tolias@kth.se}}
\abstract[Abstract]{The numerous idealizations that are involved in the exact microscopic statistical description of complex (dusty) plasmas are discussed in detail. The two prevailing approaches in the Klimontovich description of dusty plasmas are reviewed in a pedagogical manner. The continuous phase space approximation is introduced, within which the more rigorous treatment should collapse to the more heuristic treatment. The plasma Klimontovich equations are shown to be identical, but there are marked differences in the dust Klimontovich equations whose origin is analyzed in depth.}
\maketitle

\section{Introduction}\label{sec:intro}

\noindent Complex (dusty) plasmas are plasma systems that are seeded with condensed matter particulates of nanometer to micrometer size and are typically engineered in low-temperature low-pressure plasma discharges\cite{Fortov2005a}. These particulates, when embedded in plasmas, get charged by constantly collecting and emitting plasma particles and radiation\cite{deAngelis1992}. There are two theoretical aspects of complex plasmas, both consequences of the charging process, that have gathered wide attention. (1) The elementary charges that reside on the dust surface are of the order of thousand for a micron-size grain. Thus, by control of the plasma conditions and dust parameters (density, size), the average dust-dust interaction energy can become exceptionally high, which implies that the coupling parameter of the dust component can exceed unity\cite{Morfill2009a}. Together with their simple visualization, this aspect makes complex plasmas ideal laboratories for the study of strong correlations\cite{Bonitz2010,Fortov2005b,Knapek2022}. This includes liquid-like, ordered crystal and even amorphous glassy behavior as well as the transitions between these (meta)stable phases. (2) The dust grains serve as a constant sink of plasma particles and the dust charge fluctuates around its quasi-equilibrium value. The former is a manifestation of the open dissipative nature and the latter a manifestation of the non-Hamiltonian nature of complex plasmas\cite{Morfill2009a}. Therefore, even weakly coupled complex plasmas are essentially non-ideal, not in terms of the coupling strength but in the sense that they differ from standard multi-component plasmas\cite{deAngelis2006}.

The theoretical description of weakly coupled un-magnetized complex plasmas is based on the method of second quantization in phase space also known as fluctuation theory\cite{Klimontovich1958,Klimontovich1967}. The central objects of fluctuation theory are the exact microscopic phase densities of each component, each obeying their own exact Klimontovich equation\cite{Klimontovich1958,Klimontovich1967}. There are two prevalent approaches to the Klimontovich description of complex plasmas that were developed nearly simultaneously \footnote{The designations "rigorous" and "heuristic" that follow are given by the present author as a convenient means of distinction.}. A \emph{rigorous approach} developed by Schram, Zagorodny and collaborators that combines elements of the established plasma and hard sphere Klimontovich descriptions\cite{Schram2000}. In this rigorous approach, the exact microscopic phase densities are constructed based on mathematical considerations and are then differentiated in time to yield the Klimontovich equations\cite{Schram2000}. This approach has been investigated in relatively few works due to its complexity\cite{Zagorodny2000a,Zagorodny2001a}. A \emph{heuristic approach} developed by Tsytovich and de Angelis that is based exclusively on the plasma Klimontovich description\cite{Tsytovich1999a}. In this heuristic approach, the mathematical expression for the exact microscopic phase densities is circumvented and the exact Klimontovich equations are built based on physical arguments. This approach is valid under the following assumptions: the dust size is much smaller than the Debye length, the dust charge is sufficiently large and the momentum transfer due to plasma absorption is very small\cite{Tsytovich1999a}. The emerging Klimontovich equations, with the aid of the full arsenal of fluctuation theory and some additional assumptions, have been employed for the study of the dust-plasma collision integrals\cite{Tsytovich2000a,Ricci2001}, the dust-dust collision integrals\cite{Tsytovich2001a}, the dust charge fluctuations and average distribution\cite{Tsytovich2002a}, the dust and plasma hydrodynamic equations\cite{Tsytovich2004a}, the spectral densities of fluctuations\cite{Tolias2012a} and the low frequency electrostatic modes\cite{Tolias2010,Tolias2012b}. The heuristic approach has predicted fascinating phenomena such as the stochastic heating of dust particles under astrophysical conditions\cite{deAngelis2005}. The stochastic heating of dust is a general process, similar to second order Fermi acceleration, that is based on the non-conservation of energy in dust-dust interactions as a consequence of dust charge fluctuations. It has been used
to address the acceleration of small grains in the interstellar medium\cite{Ivlev2010,Hoang2012} and has also emerged from other theoretical approaches\cite{Ivlev2010,Marmolino2011}.

The rigorous approach should converge to the heuristic approach, once the validity conditions of the heuristic Klimontovich equations are converted into mathematical limits and applied to the rigorous Klimontovich equations. However, the two approaches are known to lead to different predictions concerning the dust charge distribution function\cite{Schram2003a}. This particular deviation does not constitute a crisis, because, as revealed by a comparison with discrete-time Markov descriptions of dust charging\cite{Matsoukas1995}, it is a consequence of the heuristic approach's assumption that the dust charge is sufficiently large. Nevertheless, it is rather imperative to investigate whether the two Klimontovich approaches yield different predictions also within their joint applicability range and to identify the source of the possible discrepancies (if any). This is the primary objective of the present work. The article is organized as follows. In Section \ref{sec:statisticalintro}, the exact statistical description of classical systems is briefly reviewed with an emphasis to the Klimontovich description of fully ionized plasmas. In Section \ref{sec:statisticaldust}, various idealizations of the most general complex (dusty) plasma system, that are necessary for consistency and tractability, are discussed. In Section \ref{sec:KlimontovichRigorous}, the rigorous Klimontovich description of dusty plasmas is presented with the emphasis put to the physical interpretation of the Klimontovich equations. In the same section, the continuous phase approximation which encompasses the additional approximations of the heuristic Klimontovich description is formulated and is applied to the rigorous Klimontovich description. In Section \ref{sec:KlimontovichHeuristic}, the heuristic Klimontovich description of dusty plasmas is presented and compared to the continuous phase space limit of the rigorous Klimontovich description. The article is concluded in Section \ref{sec:Outro} with a summary and an outlook.

\section{Microscopic statistical descriptions}\label{sec:statisticalintro}

\subsection{Statistical description of classical systems}

\noindent There are two equivalent treatments of the microscopic statistical description of classical $N-$body systems\cite{Nicholson1983,Schram1991,Klimontovich1995,Liboff2003,Ichimaru2018}. They essentially constitute ab initio formalisms, whose starting point is the exact dynamic evolution of the system on phase spaces of different dimensionality. As a consequence, they are of no direct use, but they allow for the development of approximate kinetic theories of varying complexity in a systematic manner. By kinetic theories, herein, we refer to approaches that investigate the evolution of the ensemble averaged one-particle distribution function.

The \emph{Liouville picture} is based on the concept of the $N-$particle distribution function $f_{N}$ that is defined on $N-$body phase space and whose evolution obeys the Liouville equation, with its general solution equivalent to the knowledge of all trajectories of the system in the $N-$phase space. Via an ingenious reformulation in terms of reduced $s$-particle distribution functions, the Liouville equation becomes equivalent to a hierarchy of $N-1$, $s$ to $s+1$ coupled, integro-differential equations, which is known as the BBGKY hierarchy and whose order becomes infinite in the thermodynamic limit\cite{Bogoliubov1946,Balescu1975}. Premature truncation of the BBGKY hierarchy at a level $s$ with an ad-hoc approximation for the level $s+1$ through the levels $1,2,...s$ (closure condition) allows the systematic derivation of different kinetic theory approximations. The Liouville picture has been applied to slightly non-ideal gases and plasmas\cite{Nicholson1983,Balescu1997}, but it is acknowledged to be more convenient for liquids\cite{McQuarrie1976,Hansen2006}. It has been generalized to quantum $N-$body systems\cite{Liboff2003,Balescu1975}, for which it is based on the concept of the $N-$particle density operator $\hat{\rho}_N$ whose evolution obeys the von-Neumann equation. In that case, reduced $s$-particle density operators lead to the quantum BBGKY hierarchy whose coordinate, momentum and Wigner representations have been well-studied in the literature\cite{Bonitz2016}.

The \emph{Klimontovich picture} is based on the concept of the microscopic phase density $f$ that is defined on the single-particle phase space and whose evolution obeys the Klimontovich equation, with its general solution equivalent to the knowledge of all trajectories of the system projected in the single particle phase space\cite{Klimontovich1958,Klimontovich1967}. Through a reformulation in terms of ensemble averaged single-time $s-$moments of the microscopic phase density that represent $s-$particle fluctuations, an $N-1$ moments hierarchy emerges together with a one-to-one correspondence with reduced s-particle distribution functions\cite{Klimontovich1967,Klimontovich1982}. In essence, the Klimontovich picture substitutes the chain of correlation functions of the Liouville picture with a chain of fluctuation moments. Similar premature truncation of the moment hierarchy allows the systematic derivation of kinetic theory approaches of varying complexity. Given the shared single particle phase space of $f$ and $f_1$, it is important to point out that $f$ is a random (stochastic) function whose ensemble average is given by the smooth continuous $f_1\equiv\Phi$ function. Extensions of the Klimontovich picture to quantum $N-$body systems had emerged nearly simultaneously with their classical counterparts but such extensions had been restricted to the Wigner representation\cite{Klimontovich1958,Klimontovich1982,Tsytovich1989}. Only very recently, an alternative generalization of the Klimontovich picture to quantum $N-$body systems has been proposed, that is based on the moments of the fluctuations of the single-particle Green's function operator on the time diagonal\cite{Schroedter2022}.

\subsection{Statistical description of classical plasmas}

\noindent The Klimontovich picture is especially suitable for slightly non-ideal plasmas\cite{Klimontovich1982,Tsytovich1989}. Collisions are essentially treated as interactions of discrete particles with the fluctuating electromagnetic fields generated by the particles themselves, which facilitates the unified treatment of the evolution of the ensemble averaged one-particle distribution function with the spectral densities of the EM field fluctuations. In this manner, the Klimontovich fluctuation theory naturally leads to the unified description of collision integrals and relaxation rates, scattering and transformation of waves, emission of radiation and particle stopping powers. Its suitability for plasmas is manifested by the elegant derivation of the Lennard-Balescu collision integral. Quoting V. N. Tsytovich from his 1989 monograph "...in the voluminous literature on this problem, which includes numerous monographs, an exceedingly cumbersome mathematical procedure is used. Today that procedure can be simplified to an extreme degree. The derivation of the Landau-Balescu collision integral takes only a few lines"\cite{Tsytovich1989}.

The Klimontovich equation of any classical system consisting of point particles can be easily derived by differentiating its known solution. In the case of a fully ionized plasma, the microscopic phase density in the single particle phase space $(\boldsymbol{r},\boldsymbol{p})$ for a charged species $\alpha=\{\mathrm{e},\mathrm{i}\}$ is simply given by the sum of $\delta-$function products centered at the exact phase space trajectories $(\boldsymbol{r}_i,\boldsymbol{p}_i)$, \emph{i.e.}\cite{Klimontovich1958,Klimontovich1967,Klimontovich1982,Tsytovich1989}
\begin{align}
f_{\alpha}(\boldsymbol{r},\boldsymbol{p};t)=\sum_{i=1}^{{N}_{\alpha}}\delta[\boldsymbol{r}-\boldsymbol{r}_i(t)]\delta[\boldsymbol{p}-\boldsymbol{p}_i(t)]\,,\label{plasmamicro}
\end{align}
where $N_{\alpha}$ is the number of particles of the species $\alpha$. In the un-magnetized non-relativistic case, the dynamics are governed by Hamilton's equations of motion for the Coulomb force. After differentiation with respect to the time, basic $\delta-$function properties together with the exact equations of motion lead to the re-emergence of $f_{\alpha}(\boldsymbol{r},\boldsymbol{p};t)$. The Klimontovich equation reads as\cite{Klimontovich1958,Klimontovich1967,Klimontovich1982,Tsytovich1989}
\begin{align}
\left[\frac{\partial}{\partial{t}}+\boldsymbol{v}\cdot\frac{\partial}{\partial\boldsymbol{r}}+e_{\alpha}\boldsymbol{E}(\boldsymbol{r};t)\cdot\frac{\partial}{\partial\boldsymbol{p}}\right]f_{\alpha}(\boldsymbol{r},\boldsymbol{p};t)=0\,,\label{plasmaKlim}
\end{align}
where $e_{\alpha}$ is the particle charge for the species $\alpha$ and where $\boldsymbol{E}(\boldsymbol{r};t)$ is the exact microscopic field. A self-consistent description necessitates that this electrostatic field is described by the Poisson equation with the particles themselves as the sources, \emph{i.e.}
\begin{align}
\nabla\cdot\boldsymbol{E}(\boldsymbol{r};t)=4\pi\sum_{\alpha}e_{\alpha}\int\,f_{\alpha}(\boldsymbol{r},\boldsymbol{p};t)\frac{d^3p}{(2\pi)^3}\,.\label{PoissonKlim}
\end{align}
Eqs.(\ref{plasmaKlim},\ref{PoissonKlim}) constitute the Klimontovich-Poisson system\cite{Klimontovich1958,Klimontovich1967,Klimontovich1982}.

\section{Statistical description of dusty plasmas}\label{sec:statisticaldust}

\noindent Plasmas seeded with particulates are genuinely complex systems, whose generic statistical description is a formidable task even at the level of the exact Klimontovich equation, \emph{i.e.} prior to the separation in ensemble averaged and fluctuating parts. Even in the case of spherical, indistinguishable (identical properties including the radius) dust grains, a large number of simplifications are necessary for the derivation of a tractable Klimontovich equation. The simplifications are necessary not only for the heuristic Klimontovich description but also for the rigorous Klimontovich description of dusty plasmas. These idealizations concern the number of phase space variables, the omission of dust surface microphysical processes, the treatment of neutrals and the handling of the non-Coulombic part of the dust interaction potential. In what follows, we shall not only discuss these idealizations in detail, but also sketch how some of these idealizations can be dropped.

\subsection{Phase space extension}

\noindent In complex (dusty) plasmas, the particulates can either be in the solid state (dust grains) or in the liquid state (droplets). In what follows, we shall refer to them as dust grains regardless of their condensed matter state. They constitute subsystems of densely packed strongly interacting atoms. In contrast to the electrons and ions which are classically point particles with constant charge over mass ratio and whose microstate in the system is defined solely by position and momentum, dust grains have an internal structure and, as a consequence, an enormous number of degrees of freedom. Nevertheless, when one considers collective phenomena in dusty plasmas, the particulates can be described without taking into account their internal structure provided that additional variables representing their microstate are defined. Such variables can be coined as mesoscopic; being microscopic when regarding the dust species as an ensemble of grains and macroscopic when regarding the atoms constituting each grain. This is a coarse-graining procedure that converts a set of many microscopic variables into a much smaller set of mesoscopic variables. Naturally, for a statistical description to be possible, each new mesoscopic variable should be accompanied with an exact dynamic equation that substitutes Hamilton's equations of motion for the disregarded microscopic variables.

The dust grains exchange particles, momentum and energy with the plasma particles, which implies that these mesoscopic variables breathe with the local plasma parameters and, therefore, can be considered as new phase space variables. Such new phase space variables are; the \emph{dust charge} $q$ that is relevant in most dusty plasma realizations\cite{Zagorodny2000a,Tsytovich1999a} and is associated with charge conservation, the \emph{dust angular momentum} $\mathbf{L}$ that could be relevant in case of strong ambient magnetic fields, inhomogeneous dust surfaces or non-spherical dust\cite{Ignatov2002a,Tsytovich2003a,Tsytovich2004b,Krasheninnikov2006a} and is associated with angular momentum conservation, the \emph{dust surface temperature} $T_{\mathrm{s}}$ that could be relevant in case of inhomogeneous plasma fluxes impinging on the grains\cite{Schram2003a,Tsytovich2004c} and is associated with energy conservation, the \emph{dust mass} $M_{\mathrm{d}}$ that could be relevant in case of strong particle absorption or intense grain vaporization\cite{Schram2003a,Ignatov2002b} and is associated with mass conservation. In the most general case, a direct consequence of the augmentation of the traditional phase space is that the microscopic dust phase densities $f_{\mathrm{d}}$ as well as the ensemble averaged one-particle dust distribution functions $\Phi_{\mathrm{d}}$ are now functions of thirteen variables $(\boldsymbol{r},\,\boldsymbol{p},\,\boldsymbol{L},\,q,\,T_{\mathrm{s}},\,M_{\mathrm{d}};\,t)$ instead of functions of seven variables $(\boldsymbol{r},\,\boldsymbol{p};\,t)$\cite{Tsytovich2004c}.

Naturally, in complex plasmas, the dust charge is the most important new phase space variable, since it dictates both the elastic Coulomb collisions with plasma particles (see scattering) and the inelastic collisions with plasma particles (see absorption). The charge extended phase space reflects one of the many non-Hamiltonian aspects of dusty plasmas\cite{Morfill2009a,Tuckerman2001}. The open dissipative nature of dusty plasmas is more evident when pointing out that the dust species also acts as an energy and particle sink\cite{Tsytovich2005a}. As we shall see, even when the classical Hamiltonian phase space is only augmented by the charge, the effect of dust on the structure of the Klimontovich equations is profound.

\subsection{Surface microphysics idealizations}

\noindent The microphysical interactions between each dust grain and its surrounding plasma can be very complex\cite{Vignitchouk2014}. A number of reasonable idealizations are necessary. Fortuitously, the idealizations that will be discussed in what follows are highly accurate for the plasma environment of low-temperature low-pressure discharges. \begin{itemize}
\item All electron emission processes from dust grains are assumed to be negligible. Even when disregarding the surface temperature phase space variable (see emission cooling), electron emission modifies the charging equation and acts as a source of electrons. It is worth pointing out that electron-induced electron emission (secondary electron emission, electron backscattering, electron reflection) and ion-induced electron emission (potential and kinetic) can be adequately described with the aid of particle incident energy and incident angle dependent yields together with electron exit energy and angular distributions\cite{Tolias2014a,Tolias2014b,Tolias2020a,Vignitchouk2018}. In addition, photoelectric emission, thermionic emission and field electron emission can also be adequately described with the aid of emitted electron flux formulas together with electron exit energy and angular distributions\cite{Gadzuk1973,Jensen2007,Jensen2019}. Thus, in principle, electron emission processes could be incorporated in the Klimontovich description without dropping the mesoscopic variables and without sacrificing the classical description. However, this would lead to an inconsistency, since the average nature of the yields, emitted fluxes and exit distributions implies that information on the inherent fluctuations that characterize these processes is already lost, see the non-integer value of the yields and the absence of energy straggling. Nevertheless, the fluctuations associated with the dust and emitted electron discreteness would still be captured.
\item All neutral atom and ion emission processes from dust grains are assumed to be negligible. Even when disregarding the surface temperature and mass phase space variables, ion emission modifies the charging equation and ion/atom emission acts as a particle source. It is worth pointing out that ion backscattering, physical sputtering and chemical sputtering can be adequately described with the aid of particle incident energy and incident angle dependent yields together with neutral particle exit energy and angular distributions\cite{Behrisch2007,Eckstein2009}. In addition, evaporation and field ion emission can also be adequately described with the aid of emitted neutral or ion flux formulas together with neutral or ion exit energy and angular distributions\cite{Muller1974}. As in the case of electron emission processes, the inclusion of atom and ion emission processes in the Klimontovich description is feasible and would partially capture the ensuing fluctuations, but is inconsistent.
\item Ion surface neutralization, also known as grain-assisted recombination, is not considered. Thus, the dust charging process is considered to proceed on the dust surface via the absorption of plasma electrons and plasma ions rather than to proceed in the {\AA}ngstr\"om vicinity of the dust surface via the absorption of plasma electrons and the emission of bound electrons that neutralize the incident plasma ions\cite{Baragiola1993}. Surface assisted recombination is an extremely effective process, especially for metallic dust due to the availability of valence electrons. It is never negligible and has a pure quantum mechanical character irrespective of the dominant charge-exchange process (resonant neutralization, Auger neutralization, quasi-resonant neutralization)\cite{Los1990,Niehus1993,Monreal2014}. Nevertheless, it should only have a weak effect on electromagnetic fluctuations.
\end{itemize}

\subsection{Ambient plasma idealizations}

\noindent The plasma is assumed to be fully ionized. Essentially, since bound electrons will not be considered at all, one is confined to a singly charged hydrogen plasma or a plasma with multi-charged heavier ions. The absence of neutrals and the complete ionization cannot be justified for the targeted plasma environment of low-temperature discharges, but they are necessary simplifications for tractability. It is apparent that a rigorous statistical theory of partially ionized complex plasmas must be developed on the basis of a quantum description\cite{Klimontovich1967b,Klimontovich1968}. However, approximate classical approaches can be followed for the atom.

In absence of ionization and recombination, the Klimontovich equation has been derived for a classical plasma-atom system containing both free and bound charged particles\cite{Klimontovich1989}. The main idea was to describe the subsystem of electrons and ions bound in pairs within the framework of a classical atomic oscillator model. For convenience, the atom-associated phase space variables concern the motion of a pair as whole as well as the motion of a particle with reduced mass in the center of mass system of the pair\cite{Klimontovich1967,Klimontovich1989}. This constitutes a phase space transformation rather than a phase space extension. It is important to stress that, while the fluctuations are exact within the model assumptions, the emerging ion-neutral and electron-neutral interactions are oversimplified.

In the presence of electron impact ionization, a heuristic Klimontovich equation has been derived for the plasma subsystem of a partially ionized plasma under some major simplifications\cite{Tsytovich2005a,Tolias2011a,deAngelis2012}. The neutrals are assumed to constitute a thermodynamic reservoir, being described with a non-fluctuating distribution function that is a Maxwell-Boltzmann distribution of a prescribed temperature. The bound electrons are not considered at all. The electron impact ionization is included as a source term in the Klimontovich equations with the ionization frequency inheriting the discreteness of the electron species. The ion-neutral and electron-neutral collisions are treated in a particle-conserving manner that can be described as the translation of the Bhatnagar-Gross-Krook operator from the ensemble averaged Boltzmann equation to the random Klimontovich equation\cite{Tsytovich2005a,Filippov2007}. For both ionization and collisions with neutrals, cross-sections need to be employed from external-to-the-theory models; this is the mark of a heuristic Klimontovich approach where consistency is sacrificed for the classical description to be retained. This treatment is limited by the disregard of atom discreteness, the implicit short range (hard-sphere like) assumption in plasma-atom interactions and the necessity for external models given the absence of a quantum description.

\subsection{Dust interaction idealizations}

\noindent In order to avoid mathematical complexities, the self consistent fine-grained interaction potential of the system is decomposed in two parts:
\begin{itemize}
\item A Coulomb non-singular part that is described by the Poisson equation for point-particle sources. As known from standard applications of the method of images in electrostatics\cite{Lekner2012}, the dust-dust and plasma-dust pair interactions are point-like only at large separations with respect to the dust size. In addition, even in the neutral dust limit, the Casimir forces between dust grains are omnipresent, emerging from the cumulative interactions between the instantaneously induced (or permanent\cite{Ramazanov2011}) multipoles arising inside the dust grains and mediated by the plasma medium\cite{Dzyaloshinskii1961,Barash1975}. The Casimir force constitutes an additional interaction that has a range of the order of few micrometers\cite{Tolias2022}, which is relatively long compared to short range forces of chemical bonding nature. Clearly, it is also negligible at large separations with respect to the micrometric dust size.
\item A singular hard-sphere part that jumps from zero to infinity at the surface of each grain. It leads to discontinuities in the charge $\&$ momentum and represents elastic contact collisions. It determines a forbidden configuration space with the particles impinging on its boundary having their phase space coordinates conveniently treated through conservation laws. In general, contact collisions between two grains may lead to sticking at low relative speeds, inelastic bouncing at intermediate speeds\cite{Ratynskaia2015}, severe plastic deformation at high speeds and fragmentation or complete vaporization at hypervelocity speeds\cite{Klinkov2005}. In principle, because sphericity is preserved at low and intermediate relative speeds, sticking and inelastic bouncing can be included in the Klimontovich description, since they are generally described by normal and tangential restitution coefficients, \emph{i.e.} ratios of the pre and post-impact velocity components of the colliding partners, which depend on the initial relative velocities, the material composition and the dust size\cite{Thornton1998,Ratynskaia2013}. In absence of spinning\cite{Thornton2009}, the tangential restitution coefficients are below unity due to frictional losses and the normal restitution coefficients are below unity due to plastic deformation as well as adhesive dissipation\cite{Tolias2017a}. It should be noted that, due to the importance of adhesion for micron-sized particulates, the dust-dust contact collisions are never elastic. However, as a crude approximation, they can be treated as instantaneous elastic collisions between infinitely hard spheres, that are inevitably accompanied by a charge redistribution. This elastic impact assumption is consistent with the disregard of multipole-multipole interactions mentioned above, which are the microscopic sources of adhesion together with the shorter range chemical bonding (covalent, metallic or ionic)\cite{Riva2017}. Finally, the redistribution of charges will depend on the conductivity of the grains and will be subject to charge conservation. The assumption of perfect electrical contact between perfectly conducting dust grains translates to an instantaneous equilibration of charges.
\end{itemize}
The above decomposition is indicative of the rigorous Klimontovich description of dusty plasmas that essentially combines the Klimontovich description of plasmas with the Klimontovich description of elastic hard spheres, as we shall see in details in the section that follows. We note that a similar decomposition of the fine-grained electrostatic potential has been performed for different reasons by Klimontovich in the statistical theory of plasma-molecular systems; into a classical Coulomb part for the charged species and a classically approximated part (of quantum mechanical origin) for the potential binding of the electron-ion pairs of the neutral species\cite{Klimontovich1989}.

\section{Rigorous Klimontovich description of dusty plasmas}\label{sec:KlimontovichRigorous}

\noindent In what follows, we shall derive the plasma and dust Klimontovich equations, for the idealized complex plasma system discussed in section \ref{sec:statisticaldust}, within the rigorous approach\cite{Schram2000}. In addition, we shall introduce our continuous phase space approximation, that is motivated by the assumptions of the heuristic Klimontovich description of dusty plasmas, and apply it to the rigorous plasma and dust Klimontovich equations. The derivations are far from straightforward and are very tedious. Thus, they will not be repeated herein, but the key mathematical steps will be emphasized. The subscript $d$ will correspond to the dust species and the subscript $\alpha=\{e,i\}$ will correspond to the plasma species (electrons or ions), $N$ will denote the particle number for each species, $m$ the particle mass and $R_{\mathrm{d}}$ the dust radius.

\subsection{The Klimontovich equation for the plasma species}\label{subsec:KlimontovichRigorousPlasma}

\noindent Within our complex plasma idealizations, the differences between the plasma Klimontovich equations in the presence and in the absence of dust originate exclusively from the annihilation of plasma particles upon contact with the surface of any grain. Hence, the $\delta$-functions positioned at the plasma particle phase space trajectories need to be multiplied by a Heaviside step function $\mathrm{H}\left(t_{i\alpha}-t\right)$, where $t_{i\alpha}$ is the absorption time of the $i$th plasma particle of the species $\alpha=\{\mathrm{e,i}\}$ that is defined as the first (in time) solution of $|\boldsymbol{r}_{i\alpha}(t)-\boldsymbol{r}_{j\mathrm{d}}(t)|=R_{\mathrm{d}},\forall1\leq{j}\leq N_{\mathrm{d}}$. In order to exclude solutions of the forbidden configuration space, the dust grain - plasma particle inelastic collision partners should approach each other from the outside, thus we should have $[\boldsymbol{r}_{i\alpha}(t_{i\alpha}^{-})-\boldsymbol{r}_{j\mathrm{d}}(t_{i\alpha}^{-})]\cdot[\boldsymbol{p}_{i\alpha}(t_{i\alpha}^{-})-\boldsymbol{p}_{j\mathrm{d}}(t_{i\alpha}^{-})]<0$. Summing over the plasma particle number, the microscopic phase density for any plasma species $\alpha$ reads as\cite{Schram2000}
\begin{equation}
f_{\alpha}(\boldsymbol{r},\boldsymbol{p};t)=\sum_{i=1}^{N_{\alpha}}\delta\left[\boldsymbol{r}-\boldsymbol{r}_{i\alpha}(t)\right]\delta\left[\boldsymbol{p}-\boldsymbol{p}_{i\alpha}(t)\right]\mathrm{H}\left(t_{i\alpha}-t\right)\,.\label{plasmamicroscopicphase}
\end{equation}
Alternatively, the microscopic plasma phase density can be formulated by replacing the temporal step function $\mathrm{H}(t_{i\alpha}-t)$ with a spatial step function $1-\sum_{j=1}^{N_{\mathrm{d}}}\mathrm{H}[R_{\mathrm{d}}-|\boldsymbol{r}-\boldsymbol{r}_{j\mathrm{d}}(t)|]$. The two formulations are equivalent in view of the properties
\begin{equation*}
\frac{\partial\mathrm{H}(t)}{\partial{t}}=\delta(t)\,,\qquad\delta\left(\left|\boldsymbol{r}_{i\alpha}(t)-\boldsymbol{r}_{j\mathrm{d}}(t)\right|-R_{\mathrm{d}}\right)=\sum_{m=1}^{m=m_{t}}\frac{\delta\left(t-t_{i\alpha,m}\right)}{\left|\left.\frac{\partial\left|\boldsymbol{r}_{i\alpha}(t)-\boldsymbol{r}_{j{\mathrm{d}}}(t)\right|}{\partial{t}}\right|_{t=t_{i\alpha,m}}\right|}\,,
\end{equation*}
where $m$ denotes the number of solutions of $|\boldsymbol{r}_{i\alpha}(t)-\boldsymbol{r}_{j\mathrm{d}}(t)|=R_{\mathrm{d}}$, that are arranged in an increasing order $t_{i\alpha}=t_{i\alpha1}<t_{i\alpha 2}<\ldots<t_{i\alpha{m}}<\ldots<t_{i\alpha{m}_{t}}$. Since only the first solution is physical and defines the annihilation time, we have
\begin{align*}
\frac{\partial\mathrm{H}\left(t_{i\alpha}-t\right)}{\partial{t}}
&=-\delta\left(t-t_{i\alpha}\right)=-\delta\left(\left|\boldsymbol{r}_{i\alpha}(t)-\boldsymbol{r}_{j\mathrm{d}}(t)\right|-R_{\mathrm{d}}\right)\left|\left.\frac{\partial\left|\boldsymbol{r}_{i\alpha}(t)-\boldsymbol{r}_{j{\mathrm{d}}}(t)\right|}{\partial{t}}\right|_{t=t_{i\alpha,1}}\right|\\
&=-\delta\left(\left|\boldsymbol{r}_{i\alpha}(t)-\boldsymbol{r}_{j\mathrm{d}}(t)\right|-R_{\mathrm{d}}\right)\left|\left.\frac{\partial\left|\boldsymbol{r}_{i\alpha}(t)-\boldsymbol{r}_{j{\mathrm{d}}}(t)\right|}{\partial[\boldsymbol{r}_{i\alpha}(t)-\boldsymbol{r}_{j{\mathrm{d}}}(t)]}\cdot\frac{\partial[\boldsymbol{r}_{i\alpha}(t)-\boldsymbol{r}_{j{\mathrm{d}}}(t)]}{\partial{t}}\right|_{t=t_{i\alpha,1}}\right|\\
&=-\delta\left(\left|\boldsymbol{r}_{i\alpha}(t)-\boldsymbol{r}_{j\mathrm{d}}(t)\right|-R_{\mathrm{d}}\right)\left|\hat{\boldsymbol{e}}_{\boldsymbol{r}_{i\alpha}(t_{i\alpha})-\boldsymbol{r}_{j\mathrm{d}}(t_{i\alpha})}\cdot\left[\boldsymbol{v}_{i\alpha}(t_{i\alpha})-\boldsymbol{v}_{j\mathrm{d}}(t_{i\alpha})\right]\right|\,.
\end{align*}
where $\hat{\boldsymbol{e}}_{\boldsymbol{x}}=\boldsymbol{x}/|\boldsymbol{x}|$ is the normalized version of the subscript vector.

Differentiation with respect to the time together with Hamilton's equations of motion for the electrostatic force, after the utilization of standard $\delta-$function properties and the above Heaviside step function property, yields the plasma Klimontovich equation\cite{Schram2000}
\begin{equation}
\left\{\frac{\partial}{\partial{t}}+\boldsymbol{v}\cdot\frac{\partial}{\partial\boldsymbol{r}}+e_{\alpha}\boldsymbol{E}(\boldsymbol{r};t)\cdot\frac{\partial}{\partial\boldsymbol{p}}\right\}f_{\alpha}(\boldsymbol{r},\boldsymbol{p};t)=\left\{\int\delta\left(\left|\boldsymbol{r}-\boldsymbol{r}^{\prime}\right|-R_{\mathrm{d}}\right)\left|\boldsymbol{v}-\boldsymbol{v}^{\prime}\right|\cos\vartheta{f}_{\mathrm{d}}\left(\boldsymbol{r}^{\prime},\boldsymbol{p}^{\prime},q;t\right)\frac{d^{3}p^{\prime}d^{3}r^{\prime}dq}{(2\pi)^{3}}\right\}f_{\alpha}(\boldsymbol{r},\boldsymbol{p};t)\,.\label{plasmaSchramfull}
\end{equation}
where $\vartheta\in[\pi/2,\pi]$ is the angle between the relative position and the relative velocity of the plasma particles and the dust grains $\cos{\vartheta}=[(\boldsymbol{r}-\boldsymbol{r}^{\prime})\cdot(\boldsymbol{v}-\boldsymbol{v}^{\prime})/|\boldsymbol{r}-\boldsymbol{r}^{\prime}||\boldsymbol{v}-\boldsymbol{v}^{\prime}|]<0$. The right hand side of the plasma Klimontovich equation is zero in the absence of dust. The right hand side term is a consequence of plasma absorption on dust and it leads to a coupling between the dust and plasma Klimontovich equations. It is inevitable that in the course of the evolution of our idealized complex plasma system, due to the continuous loss of plasma particles on the dust grains, all the plasma particles would eventually be absorbed on the dust grains. Therefore, the inclusion of a source of plasma particles is essential\cite{Tsytovich1999a}. The characteristics of the source can have deep effects on the behavior of the system as far as ensemble averaged quantities and fluctuating quantities are concerned\cite{Tsytovich2005a,Tolias2011a}. In the most general case, at the level of exact quantities, the plasma source term will be a functional of the microscopic plasma phase densities (e.g. electron impact ionization) and a functional of the microscopic dust phase density (e.g electron or ion emission), $s_{\alpha}[f_{\alpha},f_{\mathrm{d}}]$. Here, we simply add a fictitious plasma source term that does not affect the dust component in order to avoid the charge cannibalism scenario at equilibrium. Thus, we have
\begin{equation}
\left\{\frac{\partial}{\partial{t}}+\boldsymbol{v}\cdot\frac{\partial}{\partial\boldsymbol{r}}+e_{\alpha}\boldsymbol{E}(\boldsymbol{r};t)\cdot\frac{\partial}{\partial\boldsymbol{p}}\right\}f_{\alpha}(\boldsymbol{r},\boldsymbol{p};t)=s_{\alpha}(\boldsymbol{r},\boldsymbol{p};t)+\left\{\int\delta\left(\left|\boldsymbol{r}-\boldsymbol{r}^{\prime}\right|-R_{\mathrm{d}}\right)\left|\boldsymbol{v}-\boldsymbol{v}^{\prime}\right|\cos\vartheta{f}_{\mathrm{d}}\left(\boldsymbol{r}^{\prime},\boldsymbol{p}^{\prime},q;t\right)\frac{d^{3}p^{\prime}d^{3}r^{\prime}dq}{(2\pi)^{3}}\right\}f_{\alpha}(\boldsymbol{r},\boldsymbol{p};t)\,.\label{plasmaSchram}
\end{equation}

\subsection{The Klimontovich equation for the dust species}\label{subsec:KlimontovichRigorousDust}

\noindent Within our complex plasma idealizations, the differences between the dust Klimontovich equation and the plasma Klimontovich equation (in absence of dust) stem from the dust charge variability, the existence of first order discontinuities in the dust momenta and charges due to plasma absorption, the existence of first order discontinuities in the dust momenta and charges due to dust-dust contact collisions and the existence of forbidden phase space configurations due to the impenetrability of the dust grains.

During the \emph{elastic contact collisions} between perfectly conducting dust spheres, momentum conservation together with kinetic energy conservation will yield a velocity discontinuity $\delta\boldsymbol{v}_{\mathrm{d}}$ directed along the normal vector at the contact point (the tangential components will be conserved), while charge conservation together with charge equilibration will yield a charge discontinuity $\delta{q}$. Tagging the colliding dust grains with $d1,d2$, one simply has
\begin{align*}
\delta \boldsymbol{v}_{\mathrm{d1}}=-\hat{\boldsymbol{e}}_{\boldsymbol{r}_{\mathrm{d1}}-\boldsymbol{r}_{\mathrm{d2}}}\left[\hat{\boldsymbol{e}}_{\boldsymbol{r}_{\mathrm{d1}}-\boldsymbol{r}_{\mathrm{d2}}}\cdot\left(\boldsymbol{v}_{\mathrm{d1}}-\boldsymbol{v}_{\mathrm{d2}}\right)\right], \quad \delta{q}_{\mathrm{d1}}=\frac{1}{2}\left(q_{\mathrm{d2}}-q_{\mathrm{d1}}\right)\,,
\end{align*}
where $\hat{\boldsymbol{e}}_{\boldsymbol{x}}=\boldsymbol{x}/|\boldsymbol{x}|$, \emph{i.e.} the normalized version of the subscript vector. During the \emph{absorption of plasma particles} on the dust grains, that constitutes an instantaneous completely inelastic contact collision within our microphysics idealization, momentum conservation alone will yield a velocity discontinuity $\delta\boldsymbol{v}_{\mathrm{d}}$ and charge conservation alone will yield a charge discontinuity $\delta{q}$. Note that energy conservation dictates that there is a change in the internal dust energy and that mass conservation dictates that there is a change in the dust mass. Both are neglected owing to $m_{\alpha}\ll{m}_{\mathrm{d}}$. This is also consistent with the disregard of the dust mass and dust surface temperature phase space variables. Tagging the dust grain with $d$ and the plasma particle with $\alpha=\{e,i\}$, one simply has
\begin{align*}
\delta\boldsymbol{v}_{\mathrm{d}}=-\frac{m_{\alpha}}{m_{\alpha}+m_{\mathrm{d}}}\left(\boldsymbol{v}_{\mathrm{d}}-\boldsymbol{v}_{\alpha}\right), \quad \delta{q}_{\mathrm{d}}=e_{\alpha}\,.
\end{align*}

Owing to these discontinuities, the dust microscopic phase density becomes non-integrable and non-differentiable at each elastic collision instant and at each plasma particle absorption instant. As a consequence, even in the distribution sense, it can neither be differentiated with respect to time to yield the dust Klimontovich equation nor be integrated with respect to $\boldsymbol{r},\boldsymbol{p},q$ to yield the dust particle number. This problem is reminiscent of the first order discontinuities that emerge in the Liouville description of hard sphere systems\cite{Petrina1990,Petrina1998}. The remedy is to divide the time axis to segments between plasma absorption and elastic contact collision events; within such time intervals the dust trajectory will be continuous. For any given dust grain $1\leq{i}\leq{N}_{\mathrm{d}}$, the collision instants for plasma absorption are defined by the solutions of $|\boldsymbol{r}_{i\mathrm{d}}(t)-\boldsymbol{r}_{j\alpha}(t)|=R_{\mathrm{d}}\forall1\leq{j}\leq N_{\alpha}$, while the collision instants for dust contact collisions are defined by the solutions of $|\boldsymbol{r}_{i{\mathrm{d}}}(t)-\boldsymbol{r}_{k\mathrm{d}}(t)|=2R_{\mathrm{d}}\forall1\leq{k}\leq{N}_{\mathrm{d}}$ with $k\neq{i}$. Let the total number of such collisions that a dust grain $i$ undergoes in $\left(t_{0},t\right)$ be $N_{\mathrm{col},i}(t)$, where initial conditions are available for the charge-extended dust trajectories at $t_0$. The collision instants are re-arranged in increasing order $t_{1,i}<t_{2,i}<\ldots<t_{n,i}\ldots<t_{N_{\mathrm{col},i}(t)}$. The phase space trajectory of the $i$th dust grain in the interval $(t_0,t_{N_{\mathrm{col},i}(t)})$ will be given by\cite{Schram2000}
\begin{align*}
\left\{\boldsymbol{r}_i(t),\boldsymbol{p}_i(t),q_i(t)\right\}=\sum_{n=0}^{N_{\mathrm{col},i}(t)-1}\left\{\boldsymbol{r}_i^{(n)}(t),\boldsymbol{p}_i^{(n)}(t),q_i^{(n)}(t)\right\}\mathrm{H}(t-t_{n,i})\mathrm{H}(t_{n+1,i}-t)\,,
\end{align*}
with the exact phase space trajectories of any dust grain $i$ in any time interval $(t_{n,i},{t}_{n+1,i})$ given by complementing the Hamilton's equations of motion for the electrostatic force and the charge conservation with the aforementioned velocity and charge discontinuities, \emph{i.e.}\cite{Schram2000}
\begin{align*}
\boldsymbol{r}_i^{(0)}(t)&=\boldsymbol{r}_i(t_0)+\frac{1}{m_{\mathrm{d}}}\int\nolimits_{t_0}^{t}\boldsymbol{p}_i^{(0)}(t^{\prime})dt^{\prime},\,\,t_0\leq{t}\leq{t}_{1,i},\,\\
\boldsymbol{r}_i^{(n)}(t)&=\boldsymbol{r}_i^{(n-1)}(t_{n,i}^{-})+\frac{1}{m_{\mathrm{d}}}\int\nolimits_{t_{n,i}}^{t}\boldsymbol{p}_i^{(n)}(t^{\prime})dt^{\prime},\,\,t_{n,i}\leq{t}\leq{t}_{n+1,i},\,\\
\boldsymbol{p}_i^{(0)}(t)&=\boldsymbol{p}_i(t_0)+q_i(t_0)\int\nolimits_{t_0}^{t}\boldsymbol{E}[\boldsymbol{r}_i^{(0)}(t^{\prime}),t^{\prime}]dt^{\prime},\,\,t_0\leq{t}\leq{t}_{1,i},\,\\
\boldsymbol{p}_i^{(n)}(t)&=\boldsymbol{p}_i^{(n-1)}(t_{n,i}^{-})+m_{\mathrm{d}}\delta{\boldsymbol{v}}_i^{(n)}+q_i(t_{n,i}^{+})\int\nolimits_{t_{n,i}}^{t}\boldsymbol{E}[\boldsymbol{r}_i^{(n)}(t^{\prime}),t^{\prime}]dt^{\prime},\,\,t_{n,i}\leq{t}\leq{t}_{n+1,i},\,\\
q_i^{(0)}(t)&=q_i(t_0),\,\,t_0\leq{t}\leq{t}_{1,i},\,\\
q_i^{(n)}(t)&=q_i^{(n-1)}(t_{n,i}^{-})+\delta{q}_i^{(n)},\,\,t_{n,i}\leq{t}\leq{t}_{n+1,i},\,
\end{align*}
with the velocity and charge discontinuities given by\cite{Schram2000}
\begin{align*}
&\delta{\boldsymbol{v}}_i^{(n)}=
\begin{cases}
-\frac{m_{\alpha}}{m_{\alpha}+m_{\mathrm{d}}}\left[\boldsymbol{v}_{i}^{(n-1)}(t_{n,i}^{-})-\boldsymbol{v}_{j\alpha}(t_{j\alpha}^{-})\right]\qquad\qquad\qquad\qquad\qquad\qquad\qquad\,\,\,\,\,\,\,\mathrm{for\,\,plasma\,\,absorption}\,,\\
-\hat{\boldsymbol{e}}_{\boldsymbol{r}_i^{(n-1)}(t_{n,i}^{-})-\boldsymbol{r}_j^{(k-1)}(t_{k,j}^{-})}\left\{\hat{\boldsymbol{e}}_{\boldsymbol{r}_i^{(n-1)}(t_{n,i}^{-})-\boldsymbol{r}_j^{(k-1)}(t_{k,j}^{-})}\cdot\left[\boldsymbol{v}_{i}^{(n-1)}(t_{n,i}^{-})-\boldsymbol{v}_{j}^{(k-1)}(t_{k,j}^{-})\right]\right\}\,\,\,\,\mathrm{for\,\,contact\,\,collisions}\,,
\end{cases}\quad\quad\,\,\,\,\,\,\\
&\delta{q}_i^{(n)}=
\begin{cases}
e_{\alpha}\qquad\qquad\qquad\qquad\qquad\,\,\,\,\,\mathrm{for\,\,plasma\,\,absorption}\,,\\
\frac{1}{2}\left[q_{j}^{(k-1)}(t_{k,j}^{-})-q_{i}^{(n-1)}(t_{n,i}^{-})\right]\,\,\,\,\mathrm{for\,\,contact\,\,collisions}\,,
\end{cases}\quad\quad\,\,\,\,\,\,
\end{align*}
where we considered that the $n$th elastic collision of the $i$th dust particle coincides with the $k$th elastic collision of a $j$th dust particle, hence $t_{n,i}=t_{k,j}$, and that the $n$th absorption event on the $i$th dust particle coincides with the annihilation time of a $j$th plasma particle of the species $\alpha$, hence $t_{n,i}=t_{j\alpha}$. It should be noted that provided that the initial phase space configuration is physical, the phase space trajectory evolution, as described above, guarantees that the future phase space configurations will be physical, \emph{i.e.} that the impenetrability of the dust grains is always respected.

It is evident that the realization of the time segmentation with a product of two Heaviside step functions overcomes the mathematical pathologies due to the first-order discontinuities. Summing over the dust grain number, the differentiable and integrable, in the distribution sense, microscopic phase density for the dust species reads as\cite{Schram2000}
\begin{equation}
f_{\mathrm{d}}(\boldsymbol{r},\boldsymbol{p},q;t)=\sum_{i=1}^{N_{\mathrm{d}}}\sum_{n=0}^{N_{\mathrm{col},i}(t)-1}\delta\left[\boldsymbol{r}-\boldsymbol{r}_{i}(t)\right]\delta\left[\boldsymbol{p}-\boldsymbol{p}_{i}(t)\right]\delta\left[q-q_{i}(t)\right]\mathrm{H}\left(t-t_{n,i}\right)\mathrm{H}\left(t_{n+1,i}-t\right)\,.\label{dustmicroscopicphase}
\end{equation}
Differentiation with respect to the time, after the utilization of numerous $\delta-$ and step-function properties (see subsection \ref{subsec:KlimontovichRigorousPlasma}) as well as the application of the limit $t_{0}\rightarrow-\infty$ in order to dispose of a term explicitly dependent on the initial conditions, yields the dust Klimontovich equation\cite{Schram2000}
\begin{align}
&\left\{\frac{\partial}{\partial t}+\boldsymbol{v}\cdot\frac{\partial}{\partial\boldsymbol{r}}+q\boldsymbol{E}(\boldsymbol{r};t)\cdot\frac{\partial}{\partial\boldsymbol{p}}\right\}f_{\mathrm{d}}(\boldsymbol{r},\boldsymbol{p},q;t)=\nonumber\\
&\sum_{\alpha}\int\delta\left(\left|\boldsymbol{r}-\boldsymbol{r}^{\prime}\right|-R_{\mathrm{d}}\right)\left\{\cos\vartheta\left|\boldsymbol{v}-\boldsymbol{v}^{\prime}\right|f_{\mathrm{d}}(\boldsymbol{r},\boldsymbol{p},q;t)-\cos\vartheta^{\prime}\left|\boldsymbol{v}-\delta\boldsymbol{v}_{\alpha}-\boldsymbol{v}^{\prime}\right|f_{\mathrm{d}}\left(\boldsymbol{r},\boldsymbol{p}-m_{\mathrm{d}}\delta\boldsymbol{v}_{\alpha},q-e_{\alpha};t\right)\right\}f_{\alpha}\left(\boldsymbol{r}^{\prime},\boldsymbol{p}^{\prime};t\right)\frac{d^{3}p^{\prime}d^{3}r^{\prime}}{(2\pi)^{3}}+\nonumber\\
&\int\delta\left(\left|\boldsymbol{r}-\boldsymbol{r}^{\prime}\right|-2R_{\mathrm{d}}\right)\cos\vartheta\left|\boldsymbol{v}-\boldsymbol{v}^{\prime}\right|\left\{f_{\mathrm{d}}(\boldsymbol{r},\boldsymbol{p},q;t)f_{\mathrm{d}}\left(\boldsymbol{r}^{\prime},\boldsymbol{p}^{\prime},q^{\prime};t\right)-f_{\mathrm{d}}(\boldsymbol{r},\boldsymbol{p}-m_{\mathrm{d}}\delta\boldsymbol{v}_{d},q-\delta{q};t)f_{\mathrm{d}}\left(\boldsymbol{r}^{\prime},\boldsymbol{p}^{\prime}+m_{\mathrm{d}}\delta\boldsymbol{v}_{d},q^{\prime}+\delta{q};t\right)\right\}\frac{d^{3}p^{\prime}d^{3}r^{\prime}dq^{\prime}}{(2\pi)^{3}}\,.\label{dustSchramfull}
\end{align}
where $\vartheta\in[\pi/2,\pi]$ is the angle between the vectors $\boldsymbol{r}-\boldsymbol{r}^{\prime}$ and $\boldsymbol{v}-\boldsymbol{v}^{\prime}$, whereas $\vartheta^{\prime}\in[\pi/2,\pi]$ is the angle between the vectors $\boldsymbol{r}-\boldsymbol{r}^{\prime}$ and $\boldsymbol{v}-\delta \boldsymbol{v}_{\alpha}-\boldsymbol{v}^{\prime}$ (the $\vartheta,\vartheta^{\prime}$ domain stems from the fact that the dust-plasma and dust-dust collision partners need to approach each other from the outside) and where $\delta\boldsymbol{v}_{\alpha},\delta\boldsymbol{v}_{\mathrm{d}},\delta{q}$ have been introduced for brevity and denote the phase space coordinate translations of the trajectory representation for the dust momentum discontinuity during plasma absorption as well as the dust momentum and dust charge discontinuity during elastic contact collisions, \emph{i.e.},
\begin{align*}
\delta\boldsymbol{v}_{\alpha}&=-\frac{m_{\alpha}}{m_{\alpha}+m_{\mathrm{d}}}\left(\boldsymbol{v}-\boldsymbol{v}^{\prime}\right)\,,\\
\delta\boldsymbol{v}_{\mathrm{d}}&=-\hat{\boldsymbol{e}}_{\boldsymbol{r}-\boldsymbol{r}^{\prime}}\cos{\vartheta}\,,\\
\delta{q}&=\frac{1}{2}\left(q^{\prime}-q\right)\,.
\end{align*}

The right hand side of the dust Klimontovich equation is zero in the plasma Klimontovich equation (in the absence of dust). The first term on the right hand side is a consequence of plasma absorption on dust and incorporates plasma capture induced discontinuities in the dust momentum and charge. It leads to a coupling between the dust and plasma Klimontovich equations as well as to a phase space $(\boldsymbol{p},q)$ non-locality. The second term on the right hand side is a consequence of dust-dust contact collisions. It leads to a non-linearity with respect to the dust microscopic phase density as well as to a phase space $(\boldsymbol{p},q)$ non-locality. It resembles the Boltzmann collision integral for the ensemble averaged distribution function in rarified gases, $J=-\int\sigma\left(\Omega,|\boldsymbol{v}_{1}-\boldsymbol{v}_{2}|\right)|\boldsymbol{v}_{1}-\boldsymbol{v}_{2}|\left(\Phi_{1}\Phi_{2}-\Phi_{1}^{\prime}\Phi_{2}^{\prime}\right)[d\Omega{d}^{3}p_{2}/(2 \pi)^{3}]$ with the momentum arguments of the average distribution functions determined by momentum and kinetic energy conservation in elastic binary collisions\cite{Liboff2003,Balescu1975}. Such a resemblance is to be expected due to the assumption that dust contact collisions are extremely short range. The only differences are due to the addition of the local charge conservation law and the fact that the grains are not point-particles. Similar terms appear in the Klimontovich equation for a system of hard spheres\cite{Petrina1990,Petrina1998}. In the Liouville picture, they can also be derived with the use of pseudo-streaming Liouville operators, \emph{i.e.}, operators that generate physical trajectories in the Liouville phase space in which the hard spheres are not overlapping\cite{Ernst1969,Sengers1972,Ernst1981}.

For the targeted environment of low-temperature low-pressure discharges, the dust charge is typically large enough and the thermal dust velocities are typically low enough, so that dust contact collisions can be considered as extremely rare. Therefore, the second term on the right hand side of the dust Klimontovich equation can be safely neglected. This leads to the simplified \emph{dust Klimontovich equation of the rigorous approach}
\begin{align}
&\left\{\frac{\partial}{\partial t}+\boldsymbol{v}\cdot\frac{\partial}{\partial\boldsymbol{r}}+q\boldsymbol{E}(\boldsymbol{r};t)\cdot\frac{\partial}{\partial\boldsymbol{p}}\right\}f_{\mathrm{d}}(\boldsymbol{r},\boldsymbol{p},q;t)=\nonumber\\
&\sum_{\alpha}\int\delta\left(\left|\boldsymbol{r}-\boldsymbol{r}^{\prime}\right|-R_{\mathrm{d}}\right)\left\{\cos\vartheta\left|\boldsymbol{v}-\boldsymbol{v}^{\prime}\right|f_{\mathrm{d}}(\boldsymbol{r},\boldsymbol{p},q;t)-\cos\vartheta^{\prime}\left|\boldsymbol{v}-\delta\boldsymbol{v}_{\alpha}-\boldsymbol{v}^{\prime}\right|f_{\mathrm{d}}\left(\boldsymbol{r},\boldsymbol{p}-m_{\mathrm{d}}\delta\boldsymbol{v}_{\alpha},q-e_{\alpha};t\right)\right\}f_{\alpha}\left(\boldsymbol{r}^{\prime},\boldsymbol{p}^{\prime};t\right)\frac{d^{3}p^{\prime}d^{3}r^{\prime}}{(2\pi)^{3}}\,.\label{dustSchram}
\end{align}
Note that, due to the absence of dust particle sinks in our idealized complex plasma system, there is no need to introduce a dust particle source similar to the treatment of the plasma species. In general, dust particle sinks are possible in case of near-instantaneous destruction by hypervelocity impacts or ion field emission and gradual destruction by sputtering or vaporization.

\subsection{The continuous phase space approximation}\label{subsec:KlimontovichRigorousContinuousApprox}

\noindent Even when dropping dust-dust contact collisions, the Klimontovich phase space of complex plasma systems is not continuous with respect to the dust and particle positions, the dust momenta or the dust charge. As underlined in the previous subsections: (1) Due to the finite size of the impenetrable dust grains, there exist forbidden configurations in which particles overlap, $\left|\boldsymbol{r}_{j\mathrm{d}}(t)-\boldsymbol{r}_{i\alpha}(t)\right|<R_{\mathrm{d}},\forall{j}\in\left\{1,N_{\mathrm{d}}\right\},i\in\left\{1,N_{\alpha}\right\}$. (2) Due to plasma absorption, there is a jump in the momentum of the dust grains after such collisional events. (3) Due to plasma absorption, there is a jump in the charge of the dust grains after such collisional events. Due to the quantized nature of the charge, the dust charge is not a continuous phase space variable and can only change by $\pm{e}$.

The dust and plasma Klimontovich equations, Eq.(\ref{plasmaSchram},\ref{dustSchram}), can be significantly simplified by invoking the \emph{continuous phase space approximation} which smooths out these discontinuities. Within this approximation: (1) The dust particles are assumed to behave as point particles in every aspect with the exclusion of the absorption cross-sections. This also implies that dust-dust contact collisions can be dropped. This necessitates that the microscopic dust and plasma phase densities do not vary significantly within the diameter of a dust grain. As a consequence, the dust radius should be much smaller than the Debye length, $R_{\mathrm{d}}\ll\lambda_{\mathrm{D}}$. (2) Momentum transfer to the dust grains due to plasma absorption is assumed to be continuous courtesy of to $m_{\alpha}\ll{m}_{\mathrm{d}}$. (3) It is assumed that the dust charge fluctuations are much smaller than the quasi-equilibrium dust charge. Given the quantized nature of the charge and the discrete nature of the charging process, this implies that $q\gg{e}_{\alpha}$ so that $e_{\alpha}\rightarrow0$ compared to the dust charge $q$, which can now be considered to vary continuously.

Let us begin with the \emph{rigorous plasma Klimontovich equation}, Eq.(\ref{plasmaSchram}). Courtesy of the $\delta-$function and the fact that the microscopic dust phase density does not vary within a dust diameter, we have ${f}_{\mathrm{d}}\left(\boldsymbol{r}^{\prime},\boldsymbol{p}^{\prime},q;t\right)\simeq{f}_{\mathrm{d}}\left(\boldsymbol{r},\boldsymbol{p}^{\prime},q;t\right)$. Courtesy of $m_{\alpha}\ll{m}_{\mathrm{d}}$, we can also employ the approximation $|\boldsymbol{v}-\boldsymbol{v}^{\prime}|=v$ which means that $\vartheta$ is now the angle between the vectors $\boldsymbol{r}-\boldsymbol{r}^{\prime}$ and $\boldsymbol{v}$. We also set $\vartheta\to\pi-\vartheta\in[0,\pi/2]$ so that the sink term becomes more obvious. Thus,
\begin{equation*}
\left\{\frac{\partial}{\partial{t}}+\boldsymbol{v}\cdot\frac{\partial}{\partial\boldsymbol{r}}+e_{\alpha}\boldsymbol{E}(\boldsymbol{r};t)\cdot\frac{\partial}{\partial\boldsymbol{p}}\right\}f_{\alpha}(\boldsymbol{r},\boldsymbol{p};t)=s_{\alpha}(\boldsymbol{r},\boldsymbol{p};t)-\left\{\int\left[\int\nolimits_{0<\vartheta<\pi/2}\delta\left(\left|\boldsymbol{r}-\boldsymbol{r}^{\prime}\right|-R_{\mathrm{d}}\right)\cos\vartheta{d}^3r^{\prime}\right]v{f}_{\mathrm{d}}\left(\boldsymbol{r},\boldsymbol{p}^{\prime},q;t\right)\frac{d^{3}p^{\prime}dq}{(2\pi)^{3}}\right\}f_{\alpha}(\boldsymbol{r},\boldsymbol{p};t)\,.
\end{equation*}
The term in brackets represents the absorption cross-sections of the plasma particles on the dust grains (see also the units), $\sigma_{\alpha}(q,v;R_{\mathrm{d}})=\int\nolimits_{0<\vartheta<\pi/2}\delta\left(\left|\boldsymbol{r}-\boldsymbol{r}^{\prime}\right|-R_{\mathrm{d}}\right)\cos\vartheta{d}^3r^{\prime}$. The cross-sections are explicitly dependent on the dust radius and all dust size dependencies of the plasma Klimontovich equation have now been isolated. The absorption cross-sections $\sigma_{\alpha}(q,v;R_{\mathrm{d}})$ can be introduced from external models such as the orbital motion limited approach\cite{Allen1992}. The rigorous plasma Klimontovich equation within the continuous phase space approximation becomes
\begin{equation}
\left\{\frac{\partial}{\partial{t}}+\boldsymbol{v}\cdot\frac{\partial}{\partial\boldsymbol{r}}+e_{\alpha}\boldsymbol{E}(\boldsymbol{r};t)\cdot\frac{\partial}{\partial\boldsymbol{p}}\right\}f_{\alpha}(\boldsymbol{r},\boldsymbol{p};t)=s_{\alpha}(\boldsymbol{r},\boldsymbol{p};t)-\left\{\int\sigma_{\alpha}(q,v;R_{\mathrm{d}})v{f}_{\mathrm{d}}\left(\boldsymbol{r},\boldsymbol{p}^{\prime},q;t\right)\frac{d^{3}p^{\prime}dq}{(2\pi)^{3}}\right\}f_{\alpha}(\boldsymbol{r},\boldsymbol{p};t)\,,\label{plasmaSchramapprox}
\end{equation}
where the bracketed term plays the role of an exact absorption frequency of plasma particles on dust grains, which is strictly defined by $\nu_{\mathrm{d}\alpha}=\int\sigma_{\alpha}(q,v;R_{\mathrm{d}})v{f}_{\mathrm{d}}\left(\boldsymbol{r},\boldsymbol{p}^{\prime},q;t\right)[d^{3}p^{\prime}dq/{(2\pi)^{3}}]$ and constitutes the stochastic counterpart of the well known average absorption frequency $\bar{\nu}_{\mathrm{d}\alpha}=\int\sigma_{\alpha}(q,v;R_{\mathrm{d}})v{\Phi}_{\mathrm{d}}\left(\boldsymbol{r},\boldsymbol{p}^{\prime},q;t\right)[d^{3}p^{\prime}dq/{(2\pi)^{3}}]$\cite{Fortov2005a}.

Let us continue with the \emph{rigorous dust Klimontovich equation}, Eq.(\ref{dustSchram}). We shall smooth out the charge and momentum discontinuities separately. Initially, we ignore momentum transfer due to plasma absorption, which leads to $\vartheta^{\prime}\simeq\vartheta$, $\left|\boldsymbol{v}-\delta\boldsymbol{v}_{\alpha}-\boldsymbol{v}^{\prime}\right|\simeq\left|\boldsymbol{v}-\boldsymbol{v}^{\prime}\right|$, $\boldsymbol{p}-m_{\mathrm{d}}\delta\boldsymbol{v}_{\alpha}\simeq\boldsymbol{p}$. Courtesy of the $\delta-$function and the fact that the microscopic plasma phase density does not strongly vary within a dust diameter, we have ${f}_{\alpha}\left(\boldsymbol{r}^{\prime},\boldsymbol{p}^{\prime};t\right)\simeq{f}_{\alpha}\left(\boldsymbol{r},\boldsymbol{p}^{\prime};t\right)$ within the integral. Courtesy of $m_{\alpha}\ll{m}_{\mathrm{d}}$, we again employ the approximation $|\boldsymbol{v}-\boldsymbol{v}^{\prime}|=v^{\prime}$. We again set $\vartheta\to\pi-\vartheta\in[0,\pi/2]$ and isolate the plasma absorption cross-sections. The right hand side now becomes
\begin{equation*}
-\sum_{\alpha}\int{e_{\alpha}}\sigma_{\alpha}(q,v^{\prime};R_{\mathrm{d}})v^{\prime}\frac{f_{\mathrm{d}}(\boldsymbol{r},\boldsymbol{p},q;t)-f_{\mathrm{d}}\left(\boldsymbol{r},\boldsymbol{p},q-e_{\alpha};t\right)}{e_{\alpha}}f_{\alpha}\left(\boldsymbol{r},\boldsymbol{p}^{\prime};t\right)\frac{d^{3}p^{\prime}}{(2\pi)^{3}}\,.
\end{equation*}
The limit $e_{\alpha}\to0$ is implicit in the continuous phase space approximation. Thus, the partial derivative of the microscopic dust phase density with respect to the charge emerges. After some rearrangements, the right hand side now becomes
\begin{equation*}
-\left\{\sum_{\alpha}\int{e_{\alpha}}\sigma_{\alpha}(q,v^{\prime};R_{\mathrm{d}})v^{\prime}f_{\alpha}\left(\boldsymbol{r},\boldsymbol{p}^{\prime};t\right)\frac{d^{3}p^{\prime}}{(2\pi)^{3}}\right\}\frac{\partial{f}_{\mathrm{d}}(\boldsymbol{r},\boldsymbol{p},q;t)}{\partial{q}}\,.
\end{equation*}
Defining the exact plasma currents flowing to the grain $I_{\alpha}(\boldsymbol{r},q;t)=\int{e_{\alpha}}\sigma_{\alpha}(q,v^{\prime};R_{\mathrm{d}})v^{\prime}f_{\alpha}\left(\boldsymbol{r},\boldsymbol{p}^{\prime};t\right)[{d^{3}p^{\prime}}/{(2\pi)^{3}}]$ as the stochastic counterpart of the well known average charging plasma currents $\bar{I}_{\alpha}(\boldsymbol{r},q;t)=\int{e_{\alpha}}\sigma_{\alpha}(q,v^{\prime};R_{\mathrm{d}})v^{\prime}\Phi_{\alpha}\left(\boldsymbol{r},\boldsymbol{p}^{\prime};t\right)[{d^{3}p^{\prime}}/{(2\pi)^{3}}]$\cite{Fortov2005a}, it is obvious that the bracketed term is the exact total plasma current flowing to the grain $I(\boldsymbol{r},q;t)=\sum_{\alpha}I_{\alpha}(\boldsymbol{r},q;t)$. Ultimately, the right hand side is simply written as
\begin{equation*}
-I(\boldsymbol{r},q;t)\frac{\partial{f}_{\mathrm{d}}(\boldsymbol{r},\boldsymbol{p},q;t)}{\partial{q}}\,.
\end{equation*}
Now we ignore charge transfer due to plasma absorption, which implies that $q-e_{\alpha}\simeq{q}$ and we employ ${f}_{\alpha}\left(\boldsymbol{r}^{\prime},\boldsymbol{p}^{\prime};t\right)\simeq{f}_{\alpha}\left(\boldsymbol{r},\boldsymbol{p}^{\prime};t\right)$ within the integral. We use the approximation $\delta\boldsymbol{v}_{\alpha}=-[m_{\alpha}/(m_{\alpha}+m_{\mathrm{d}})](\boldsymbol{v}-\boldsymbol{v}^{\prime})\simeq(m_{\alpha}/m_{\mathrm{d}})\boldsymbol{v}^{\prime}$ given that $v^{\prime}\gg{v}$ and $m_{\mathrm{d}}\gg{m}_{\alpha}$. The right hand side now becomes
\begin{equation*}
\sum_{\alpha}\int\delta\left(\left|\boldsymbol{r}-\boldsymbol{r}^{\prime}\right|-R_{\mathrm{d}}\right)\frac{m_{\alpha}}{m_{\mathrm{d}}}\boldsymbol{v}^{\prime}\cdot\frac{\cos\vartheta\left|\boldsymbol{v}-\boldsymbol{v}^{\prime}\right|f_{\mathrm{d}}(\boldsymbol{r},\boldsymbol{p},q;t)-\cos\vartheta^{\prime}\left|\boldsymbol{v}-\boldsymbol{v}^{\prime}-\frac{m_{\alpha}}{m_{\mathrm{d}}}\boldsymbol{v}^{\prime}\right|f_{\mathrm{d}}\left(\boldsymbol{r},\boldsymbol{p}-m_{\mathrm{d}}\frac{m_{\alpha}}{m_{\mathrm{d}}}\boldsymbol{v}^{\prime},q;t\right)}{\frac{m_{\alpha}}{m_{\mathrm{d}}}\boldsymbol{v}^{\prime}}f_{\alpha}\left(\boldsymbol{r},\boldsymbol{p}^{\prime};t\right)\frac{d^{3}p^{\prime}d^{3}r^{\prime}}{(2\pi)^{3}}\,.
\end{equation*}
The limit $(m_{\alpha}/m_{\mathrm{d}})\boldsymbol{v}^{\prime}\to0$ is implicit in the continuous phase space approximation. Thus, the partial derivative of the microscopic dust phase density with respect to the velocity emerges. After some rearrangements, the right hand side now becomes
\begin{equation*}
\frac{\partial}{\partial\boldsymbol{p}}\cdot\sum_{\alpha}\int\delta\left(\left|\boldsymbol{r}-\boldsymbol{r}^{\prime}\right|-R_{\mathrm{d}}\right)m_{\alpha}\boldsymbol{v}^{\prime}\cos\vartheta\left|\boldsymbol{v}-\boldsymbol{v}^{\prime}\right|f_{\mathrm{d}}(\boldsymbol{r},\boldsymbol{p},q;t)f_{\alpha}\left(\boldsymbol{r},\boldsymbol{p}^{\prime};t\right)\frac{d^{3}p^{\prime}d^{3}r^{\prime}}{(2\pi)^{3}}\,.
\end{equation*}
We again set $\vartheta\to\pi-\vartheta\in[0,\pi/2]$, employ $v^{\prime}\gg{v}$ and isolate the plasma absorption cross-sections. After some operator interchanges, the right hand side becomes
\begin{equation*}
\left\{-\sum_{\alpha}\int{m}_{\alpha}\boldsymbol{v}^{\prime}\sigma_{\alpha}(q,v^{\prime};R_{\mathrm{d}}){v}^{\prime}f_{\alpha}\left(\boldsymbol{r},\boldsymbol{p}^{\prime};t\right)\frac{d^{3}p^{\prime}}{(2\pi)^{3}}\right\}\cdot\frac{\partial{f_{\mathrm{d}}(\boldsymbol{r},\boldsymbol{p},q;t)}}{\partial\boldsymbol{p}}\,.
\end{equation*}
Defining the exact plasma absorption forces $\boldsymbol{F}_{\mathrm{b,\alpha}}(\boldsymbol{r},q;t)=\int{m}_{\alpha}\boldsymbol{v}^{\prime}\sigma_{\alpha}(q,v^{\prime};R_{\mathrm{d}}){v}^{\prime}f_{\alpha}\left(\boldsymbol{r},\boldsymbol{p}^{\prime};t\right)[{d^{3}p^{\prime}}/{(2\pi)^{3}}]$ as the stochastic counterparts of the known average forces due to plasma absorption $\bar{\boldsymbol{F}}_{\mathrm{b,\alpha}}(\boldsymbol{r},q;t)=\int{m}_{\alpha}\boldsymbol{v}^{\prime}\sigma_{\alpha}(q,v^{\prime};R_{\mathrm{d}}){v}^{\prime}\Phi_{\alpha}\left(\boldsymbol{r},\boldsymbol{p}^{\prime};t\right)[{d^{3}p^{\prime}}/{(2\pi)^{3}}]$\cite{Fortov2005a}, it is clear that the bracketed term is the microscopic bombardment force due to plasma absorption $\boldsymbol{F}_{\mathrm{b}}(\boldsymbol{r},q;t)=\sum_{\alpha}\boldsymbol{F}_{\mathrm{b,\alpha}}(\boldsymbol{r},q;t)$. Ultimately, the right hand side is simply written as
\begin{equation*}
-\boldsymbol{F}_{\mathrm{b}}(\boldsymbol{r},q;t)\cdot\frac{\partial{f_{\mathrm{d}}(\boldsymbol{r},\boldsymbol{p},q;t)}}{\partial\boldsymbol{p}}\,.
\end{equation*}
Combining the above, the rigorous dust Klimontovich equation within the continuous phase space approximation becomes
\begin{equation}
\left\{\frac{\partial}{\partial{t}}+\boldsymbol{v}\cdot\frac{\partial}{\partial\boldsymbol{r}}+\left[q\boldsymbol{E}(\boldsymbol{r};t)+\boldsymbol{F}_{\mathrm{b}}(\boldsymbol{r},q;t)\right]\cdot\frac{\partial}{\partial\boldsymbol{p}}+I(\boldsymbol{r},q;t)\frac{\partial}{\partial{q}}\right\}f_{\mathrm{d}}(\boldsymbol{r},\boldsymbol{p},q;t)=0\,.\label{dustSchramapprox}
\end{equation}
It is important to emphasize that the above treatment is equivalent to the Taylor expansion of the term of the integrand $\cos\vartheta^{\prime}\left|\boldsymbol{v}-\delta\boldsymbol{v}_{\alpha}-\boldsymbol{v}^{\prime}\right|f_{\mathrm{d}}\left(\boldsymbol{r},\boldsymbol{p}-m_{\mathrm{d}}\delta\boldsymbol{v}_{\alpha},q-e_{\alpha};t\right)$ of the rigorous dust Klimontovich equation, see Eq.(\ref{dustSchram}), simultaneously with respect to the two small parameters $e_{\alpha}$, $(m_{\alpha}/m_{\mathrm{d}})\boldsymbol{v}^{\prime}$ and to retaining the two first order terms. The equivalent separate evaluation of the two terms was preferred for the purposes of presentation. A Taylor expansion up to the three second order terms was performed in Ref.\cite{Schram2000} and the first order form was also provided, but the authors did not separate the plasma absorption cross-sections. Note also that even higher order Taylor expansions are rather straightforward to obtain.

It is also important to point out that this dust Klimontovich equation has a convective derivative form $Df_{\mathrm{d}}(\boldsymbol{r},\boldsymbol{p},q;t)/Dt=0$ in the charge extended phase-space $\{\boldsymbol{r},\boldsymbol{p},q\}$ but does not have a continuity form courtesy of $\partial{I}(q,\boldsymbol{r};t)/\partial{q}\neq0$. The convective derivative form implies that this dust Klimontovich equation could have been derived in a much simpler manner. In particular, as suggested by its name, the continuous phase space approximation implies that there are no phase space discontinuities. Combined with the absence of sinks or sources of dust particles, the dust microscopic phase density in the charge extended phase space, simply reads as
\begin{equation*}
f_{\mathrm{d}}(\boldsymbol{r},\boldsymbol{p},q;t)=\sum_{i=1}^{N_{\mathrm{d}}}\delta\left[\boldsymbol{r}-\boldsymbol{r}_{i}(t)\right]\delta\left[\boldsymbol{p}-\boldsymbol{p}_{i}(t)\right]\delta\left[q-q_{i}(t)\right]\,.
\end{equation*}
where the dust phase space trajectories are determined by
\begin{align*}
\frac{\partial\boldsymbol{r}_{i}(t)}{\partial{t}}&=\boldsymbol{v}_i(t)\,,\\
\frac{\partial\boldsymbol{p}_{i}(t)}{\partial{t}}&=q_i(t)\boldsymbol{E}[\boldsymbol{r}_{i}(t);(t)]+\boldsymbol{F}_{\mathrm{b}}[\boldsymbol{r}_i(t),q_i(t);t]\,,\\
\frac{\partial\boldsymbol{q}_{i}(t)}{\partial{t}}&=I[\boldsymbol{r}_i(t),q_i(t);t]\,.
\end{align*}
Direct differentiation of the above dust microscopic phase density with respect to the time together with basic $\delta-$function properties and the exact dynamic equations for the phase space trajectories mentioned above lead to the re-emergence of $f_{\mathrm{d}}(\boldsymbol{r},\boldsymbol{p};t)$ and the rigorous dust Klimontovich equation within the continuous phase approximation, Eq.(\ref{dustSchramapprox}). This simplification is facilitated by the use of external models for the absorption cross-sections which enter the microscopic charging current $I$ and the microscopic bombardment force $\boldsymbol{F}_{\mathrm{b}}$. Despite the much more simplified derivation, the correct procedure is derive the rigorous dust Klimontovich equation from the general microscopic dust phase densities and then apply the continuous phase space approximation rather than to apply the continuous phase space approximation to the microscopic dust phase densities and then derive the dust Klimontovich equation. In the present case the result is indeed the same, but the interchange of the complicated limit implied by the continuous phase space approximation with the integro-differential operators involved in the Klimontovich equation derivation is not guaranteed.

\section{Heuristic Klimontovich description of dusty plasmas}\label{sec:KlimontovichHeuristic}

\subsection{The Klimontovich equation for the plasma and dust species}\label{subsec:KlimontovichHeuristicPlasmaDust}

\noindent In what follows, we shall construct the plasma and dust Klimontovich equations, for the idealized complex plasma system discussed in section \ref{sec:statisticaldust}, within the heuristic approach\cite{Tsytovich1999a}. The basic assumptions of the Tsytovich and de Angelis treatment that allow to simplify the mathematical procedure are equivalent to the continuous phase space approximation introduced in subsection \ref{subsec:KlimontovichRigorousContinuousApprox}. The plasma and dust Klimontovich equations are not strictly derived but rather constructed on the basis of physical arguments centered around the conservation or non-conservation of the continuous phase space points.

\emph{For the plasma particles}, simple phase space point conservation arguments together with the physical meaning of the plasma absorption cross-sections on the dust grains directly lead to the plasma Klimontovich equation\cite{Tsytovich1999a}
\begin{align}
\left\{\frac{\partial}{\partial{t}}+\boldsymbol{v}\cdot\frac{\partial}{\partial\boldsymbol{r}}+e_{\alpha}\boldsymbol{E}(\boldsymbol{r};t)\cdot\frac{\partial}{\partial\boldsymbol{p}}\right\}f_{\alpha}(\boldsymbol{r},\boldsymbol{p};t)=s_{\alpha}(\boldsymbol{r},\boldsymbol{p};t)-\left\{\int\sigma_{\alpha}(q,v;R_{\mathrm{d}})v{f}_{\mathrm{d}}\left(\boldsymbol{r},\boldsymbol{p}^{\prime},q;t\right)\frac{d^{3}p^{\prime}dq}{(2\pi)^{3}}\right\}f_{\alpha}(\boldsymbol{r},\boldsymbol{p};t)\,,\label{plasmaTsytovich}
\end{align}

\emph{For the dust grains}, the Liouville picture is considered. The phase space is assumed to be continuous and charge extended. Therefore, the $7N_{\mathrm{d}}-$dimensional phase space can be compactly represented by $\boldsymbol{x}=\{\boldsymbol{x}_1,\boldsymbol{x}_2,\ldots,\boldsymbol{x}_{N_{\mathrm{d}}}\}$, where $\boldsymbol{x}_i=\{\boldsymbol{r}_i,\boldsymbol{p}_i,q_i\}$. An equivalent representation is $\boldsymbol{x}=\{\boldsymbol{r}^{N_{\mathrm{d}}},\boldsymbol{p}^{N_{\mathrm{d}}},q^{N_{\mathrm{d}}}\}$ with $\boldsymbol{r}^{N_{\mathrm{d}}}=\{\boldsymbol{r}_1,\boldsymbol{r}_2,\ldots,\boldsymbol{r}_{N_{\mathrm{d}}}\}$, $\boldsymbol{p}^{N_{\mathrm{d}}}=\{\boldsymbol{p}_1,\boldsymbol{p}_2,\ldots,\boldsymbol{p}_{N_{\mathrm{d}}}\}$, $\boldsymbol{q}^{N_{\mathrm{d}}}=\{q_1,q_2,\ldots,q_{N_{\mathrm{d}}}\}$. As in the fluid dynamics derivation of the classical Liouville equation\cite{Nicholson1983,Liboff2003,Balescu1975}, it is assumed that phase space points are neither created nor annihilated, thus the rate of charge of phase space points in a given volume is equal to the net flux of points that pass through the closed surface that encloses this given volume. This implies that the $f_{N,\mathrm{d}}$ evolution is described by\cite{Tsytovich1999a}
\begin{align*}
&\frac{\partial{f}_{N,\mathrm{d}}(\boldsymbol{x};t)}{\partial{t}}+\nabla_{\boldsymbol{x}}\cdot\left[\dot{\boldsymbol{x}}{f}_{N,\mathrm{d}}(\boldsymbol{x};t)\right]=0\Rightarrow\\
&\frac{\partial{f}_{N,\mathrm{d}}(\boldsymbol{x};t)}{\partial{t}}+\nabla_{\boldsymbol{r}^{N_{\mathrm{d}}}}\cdot\left[\dot{\boldsymbol{r}}^{N_{\mathrm{d}}}{f}_{N,\mathrm{d}}(\boldsymbol{x};t)\right]+\nabla_{\boldsymbol{p}^{N_{\mathrm{d}}}}\cdot\left[\dot{\boldsymbol{p}}^{N_{\mathrm{d}}}{f}_{N,\mathrm{d}}(\boldsymbol{x};t)\right]+\nabla_{\boldsymbol{q}^{N_{\mathrm{d}}}}\cdot\left[\dot{\boldsymbol{q}}^{N_{\mathrm{d}}}{f}_{N,\mathrm{d}}(\boldsymbol{x};t)\right]=0\Rightarrow\\
&\frac{\partial{f}_{N,\mathrm{d}}(\boldsymbol{x};t)}{\partial{t}}+\sum_{i=1}^{N_{\mathrm{d}}}\frac{\partial}{\partial\boldsymbol{r}_i}\cdot\left[\dot{\boldsymbol{r}}_i{f}_{N,\mathrm{d}}(\boldsymbol{x};t)\right]+\sum_{i=1}^{N_{\mathrm{d}}}\frac{\partial}{\partial\boldsymbol{p}_i}\cdot\left[\dot{\boldsymbol{p}}_i{f}_{N,\mathrm{d}}(\boldsymbol{x};t)\right]+\sum_{i=1}^{N_{\mathrm{d}}}\frac{\partial}{\partial{q}_i}\left[\dot{q}_i{f}_{N,\mathrm{d}}(\boldsymbol{x};t)\right]=0\,.
\end{align*}
Integrating over the phase space variables (positions, momenta, charges) of all particles but one, \emph{i.e.} $\{\boldsymbol{x}_2,\ldots,\boldsymbol{x}_{N_{\mathrm{d}}}\}$; all terms that concern the integration phase space variables vanish, while in the surviving terms the microscopic dust phase density substitutes the $N$-particle distribution function. Employing the notation $\{\boldsymbol{r}_1,\boldsymbol{p}_1,q_1\}=\{\boldsymbol{r},\boldsymbol{p},q\}$, one obtains
\begin{align*}
&\frac{\partial{f}_{\mathrm{d}}(\boldsymbol{r},\boldsymbol{p},q;t)}{\partial{t}}+\frac{\partial}{\partial\boldsymbol{r}}\cdot\left[\dot{\boldsymbol{r}}f_{\mathrm{d}}(\boldsymbol{r},\boldsymbol{p},q;t)\right]+\frac{\partial}{\partial\boldsymbol{p}}\cdot\left[\dot{\boldsymbol{p}}f_{\mathrm{d}}(\boldsymbol{r},\boldsymbol{p},q;t)\right]+\frac{\partial}{\partial{q}}\left[\dot{q}f_{\mathrm{d}}(\boldsymbol{r},\boldsymbol{p},q;t)\right]=0\,.
\end{align*}
Utilizing the dynamic equations $\dot{\boldsymbol{r}}=\boldsymbol{v}$, $\dot{\boldsymbol{p}}=q\boldsymbol{E}(\boldsymbol{r};t)$, $\dot{q}=I(\boldsymbol{r},q;t)$, one has
\begin{align*}
&\frac{\partial{f}_{\mathrm{d}}(\boldsymbol{r},\boldsymbol{p},q;t)}{\partial{t}}+\frac{\partial}{\partial\boldsymbol{r}}\cdot\left[\boldsymbol{v}f_{\mathrm{d}}(\boldsymbol{r},\boldsymbol{p},q;t)\right]+\frac{\partial}{\partial\boldsymbol{p}}\cdot\left[q\boldsymbol{E}(\boldsymbol{r};t)f_{\mathrm{d}}(\boldsymbol{r},\boldsymbol{p},q;t)\right]+\frac{\partial}{\partial{q}}\left[I(\boldsymbol{r},q;t)f_{\mathrm{d}}(\boldsymbol{r},\boldsymbol{p},q;t)\right]=0\,.
\end{align*}
The independence of the phase space variables $\{\boldsymbol{r},\boldsymbol{p},q\}$ leads to
\begin{align*}
&\left\{\frac{\partial}{\partial{t}}+\boldsymbol{v}\cdot\frac{\partial}{\partial\boldsymbol{r}}+q\boldsymbol{E}(\boldsymbol{r};t)\cdot\frac{\partial}{\partial\boldsymbol{p}}\right\}{f}_{\mathrm{d}}(\boldsymbol{r},\boldsymbol{p},q;t)+\frac{\partial}{\partial{q}}\left[I(\boldsymbol{r},q;t)f_{\mathrm{d}}(\boldsymbol{r},\boldsymbol{p},q;t)\right]=0\,.
\end{align*}
Ultimately, this is straightforwardly rewritten as\cite{Tsytovich1999a}
\begin{align}
&\left\{\frac{\partial}{\partial{t}}+\boldsymbol{v}\cdot\frac{\partial}{\partial\boldsymbol{r}}+q\boldsymbol{E}(\boldsymbol{r};t)\cdot\frac{\partial}{\partial\boldsymbol{p}}+I(\boldsymbol{r},q;t)\frac{\partial}{\partial{q}}\right\}{f}_{\mathrm{d}}(\boldsymbol{r},\boldsymbol{p},q;t)=-\frac{\partial{I}(\boldsymbol{r},q;t)}{\partial{q}}f_{\mathrm{d}}(\boldsymbol{r},\boldsymbol{p},q;t)\,.\label{dustTsytovich}
\end{align}

\subsection{Comparison}\label{subsec:KlimontovichHeuristicRigorous}

\noindent We are in the position to compare the continuous phase space limit of the rigorous Klimontovich description of dusty plasmas with the heuristic Klimontovich description of dusty plasmas. It is evident that the plasma Klimontovich equations are identical, compare Eq.(\ref{plasmaSchramapprox}) to Eq.(\ref{plasmaTsytovich}). On the other hand, there are two differences in the dust Klimontovich equations, compare Eq.(\ref{dustSchramapprox}) to Eq.(\ref{dustTsytovich}). (1) The microscopic bombardment force due to the continuous plasma absorption $\boldsymbol{F}_{\mathrm{b}}(\boldsymbol{r},q;t)$ is not taken into account in the heuristic approach. This is an unacknowledged omission of the heuristic approach that stems from the dynamic equation $\dot{\boldsymbol{p}}=q\boldsymbol{E}(\boldsymbol{r};t)$. It can be simply remedied by employing the correct dynamic equation $\dot{\boldsymbol{p}}=q\boldsymbol{E}(\boldsymbol{r};t)+\boldsymbol{F}_{\mathrm{b}}(\boldsymbol{r},q;t)$. (2) The non-convective term $-[\partial{I}(\boldsymbol{r},q;t)/\partial{q}]f_{\mathrm{d}}(\boldsymbol{r},\boldsymbol{p},q;t)$ of the heuristic approach is not present in the continuous phase space limit of the rigorous approach. This directly originates from the assumption of phase space point conservation in the $7N_{\mathrm{d}}-$dimensional phase space of the Liouville picture that is invoked in the heuristic approach. The heuristic and rigorous approaches become identical within the continuous phase space approximation only when $\partial{I}(\boldsymbol{r},q;t)/\partial{q}\equiv0$. This cannot be true, since we have ${I}(\boldsymbol{r},q;t)=\sum_{\alpha}\int{e_{\alpha}}\sigma_{\alpha}(q,v^{\prime};R_{\mathrm{d}})v^{\prime}f_{\alpha}\left(\boldsymbol{r},\boldsymbol{p}^{\prime};t\right)[{d^{3}p^{\prime}}/{(2\pi)^{3}}]$ and $\partial\sigma_{\alpha}(q,v^{\prime};R_{\mathrm{d}})/\partial{q}\neq0$. This is obvious from the plasma absorption cross sections of the
orbital motion limited approach\cite{Fortov2005a,Allen1992}, an approach which is consistent with the assumption $R_{\mathrm{d}}\ll\lambda_{\mathrm{D}}$ of the continuous phase space approximation, that read as
\begin{equation*}
\sigma_{\alpha}(q,v;R_{\mathrm{d}})=\pi{R}_{\mathrm{d}}^2\left(1-\frac{2qe_{\alpha}}{R_{\mathrm{d}}m_{\alpha}v^2}\right)\mathrm{H}\left(\frac{1}{2}m_{\alpha}v^2-\frac{qe_{\alpha}}{R_{\mathrm{d}}}\right)\,.
\end{equation*}
At this point, it should be emphasized that the continuous phase space approximation of the rigorous Klimontovich description yields $\sigma_{\alpha}(q,v;R_{\mathrm{d}})=\int\nolimits_{0<\vartheta<\pi/2}\delta\left(\left|\boldsymbol{r}-\boldsymbol{r}^{\prime}\right|-R_{\mathrm{d}}\right)\cos\vartheta{d}^3r^{\prime}$. Thus, only the plasma velocity and dust radius dependencies of the cross-sections are obvious, $\sigma_{\alpha}(v;R_{\mathrm{d}})$. The charge dependence has been included, $\sigma_{\alpha}(q,v;R_{\mathrm{d}})$, because it needs to be there from general physics arguments and in view of the orbital motion limited cross-sections. It is also important to point out that the use of the above cross-sections is strictly permissible only at low frequencies that are characteristic of dust motion, for which the plasma discreteness - see natural plasma fluctuations\cite{Tsytovich1995,Tolias2015} - can be assumed to be negligible\cite{Tsytovich1999a,Tsytovich2002a}.

\section{Summary and outlook}\label{sec:Outro}

\noindent The statistical description of complex (dusty) plasmas at the exact level of the microscopic phase densities and of the accompanying Klimontovich equations is a notoriously difficult task, even for sufficiently but not extensively idealized versions of these systems. An available rigorous approach is analyzed that constructs the plasma and dust microscopic phase densities based on the exact phase space trajectories and derives the dust and plasma Klimontovich equations after a time differentiation based on mathematical considerations. An available heuristic approach is also analyzed that constructs the dust and plasma Klimontovich equations directly based on phase space point conservation arguments and additional physical assumptions. A continuous phase space approximation is introduced that comprises of three small parameters ($R_{\mathrm{d}}\ll\lambda_{\mathrm{D}}$, $e_{\alpha}\ll{q}$, $m_{\alpha}\ll{m}_{\mathrm{d}}$) which formalizes all the assumptions of the heuristic approach that refer to the Klimontovich level. The continuous phase space approximation is applied to the rigorous plasma and dust Klimontovich equations, which become considerably simplified. It is proven that the same Klimontovich equations emerge when the continuous phase space approximation is applied to the rigorous microscopic phase densities that are then differentiated in time.

The term-by-term comparison between the continuous phase space limit of the rigorous Klimontovich equations and the heuristic Klimontovich equations revealed that the plasma Klimontovich equations are identical but that there are discrepancies between the dust Klimontovich equations. The first discrepancy concerns the omission of the microscopic plasma bombardment force in the heuristic approach and the second discrepancy concerns an additional spurious term in the heuristic approach that ultimately stems from the charge dependence of the plasma absorption cross-sections on dust. Given the rather extensive application of the heuristic approach to many weakly coupled (gaseous) dusty plasma kinetic and fluctuation studies due to its relative simplicity, it is important that such investigations are repeated with the formally more correct and nearly equally complex continuous phase space limit of the rigorous approach. This would lead to more exact results and would allow us to quantify the physical impact of the aforementioned two discrepancies in different low-temperature plasma discharge environments.

\bibliography{klim_bib}

\end{document}